\documentclass[%
twocolumn,%
amsmath,%
amssymb
aps,%
twocolumn,%
superscriptaddress,%
byrevtex,%
showpacs
]%
{revtex4}

\usepackage{graphicx}%      Include figure files
\usepackage{dcolumn}%       Align table columns on decimal point
\usepackage{bm}%            bold math
\usepackage{epsfig}
\usepackage{color}
\usepackage{subfigure}

\newcommand{\tmfloatsmall}[2]{
\begin{figure}
#1
\caption{#2}
\end{figure}}

\newcommand{\tmfloatsmallfirst}[2]{
\begin{figure}[b]
#1
\caption{#2}
\end{figure}}
\newcommand{\tmfloatsmalltop}[2]{
\begin{figure}[t]
#1
\caption{#2}
\end{figure}}

\begin{document}

\title{Quantum transport through STM-lifted single PTCDA molecules}

\author{Florian Pump}
\affiliation{Institute for Materials Science and Max Bergmann
Center of Biomaterials, Dresden University of Technology, D-01062
Dresden, Germany}
\author{Ruslan Temirov}
\affiliation{Institut f{\"u}r Bio-und Nanosysteme 3, JARA,
Forschungszentrum J{\"u}lich, D-52425 J{\"u}lich, Germany}
\author{Olga Neucheva}
\affiliation{Institut f{\"u}r Bio-und Nanosysteme 3, JARA,
Forschungszentrum J{\"u}lich, D-52425 J{\"u}lich, Germany}
\author{Serguei Soubatch}
\affiliation{Institut f{\"u}r Bio-und Nanosysteme 3, JARA,
Forschungszentrum J{\"u}lich, D-52425 J{\"u}lich, Germany}
\author{Stefan Tautz}
\affiliation{Institut f{\"u}r Bio-und Nanosysteme 3, JARA,
Forschungszentrum J{\"u}lich, D-52425 J{\"u}lich, Germany}
\author{Michael Rohlfing}
\affiliation{Department of Physics, University of Osnabr{\"u}ck,
D-49069 Osnabr{\"u}ck, Germany}
\author{Gianaurelio Cuniberti}
\affiliation{Institute for Materials Science and Max Bergmann
Center of Biomaterials, Dresden University of Technology,
D-01062 Dresden, Germany}
\date{\today}

\begin{abstract}
Using a scanning tunneling microscope we have measured the quantum
conductance through a PTCDA molecule for different configurations
of the tip-molecule-surface junction. A peculiar conductance
resonance arises at the Fermi level for certain tip to surface
distances. We have relaxed the molecular junction coordinates and
calculated transport by means of the Landauer/Keldysh approach.
The zero bias transmission calculated for fixed tip positions in
lateral dimensions but different tip substrate distances show a
clear shift and sharpening of the molecular chemisorption level
on increasing the STM-surface distance, in agreement with
experiment.
\end{abstract}


\maketitle

\section{Introduction}\label{sect:Intro}
Since the emergence of molecular electronics in the late nineties of
the previous century enormous progress has been achieved in both 
experimental and theoretical understanding of the underlying
microscopic mechanisms governing the charge (and also recently spin)
migration at the molecular scale~\cite{:Cun2005}. In this sense,
the realization of electronic functions at nanometer scales, as
pioneered by the visionary work of Aviram and Ratner in
1973~\cite{AR74a}, is slowly turning from a dream into a reality.
However, there are still formidable challenges to be overcome.

The current state of affairs can be characterized by the following
observations: in nanometer-sized junctions the connection between
the molecule and the electrode greatly affects the current-voltage
characteristics \cite{nitzan03}, but there are currently no robust
methods to image and determine the precise adsorption site and
conformation of the molecule on this length scale \cite{:Joa2005}.
We are thus facing a dilemma: on the one hand, we know from theory
that contacts are extremely important \cite{:Val2007, :Ke2005, :Kau2008}, on the other hand there is
-- within conventional approaches -- no way to obtain atomic-scale
information on these contacts. The statistical analysis of large
data sets (thousands of experiments) has proven to be one way out
of the dilemma. However, this procedure is evidently very
laborious. For this reason, we wish to explore an alternative
approach in the present paper, namely to chemically contact single
molecules in a reproducible way, such that reproducible transport
spectra can be obtained even in small data sets. This approach is
currently being explored; see e.g. refs.~\cite{:Tem2008, :Nee2007}. It allows for transport experiments in structurally
well-controlled junctions and thus offers an excellent interface
to transport calculations. Contacts to single atoms have
also been {studied}~\cite{:Nee2007a, :Kro2007, :Lim2005,
:Yaz1996}.

Experimentally, there are three main techniques which form the
pillars of the current state-of-the-art: mechanically controllable
break junctions
(MCBJ)~\cite{:Dre2005,:Sch97,:Sch98,:Elb2005,:Rei2002,:Rei2003,:Web2001,:Xie2004,:Smi2002,:Agr2003},
electromigration techniques (EM)~\cite{EM1}, and scanning tunnelling
microscopy
(STM)~\cite{:Tem2008,:Rep2005,:Sti98,:Sti99,:Cho2006,:Yan2002,:Bra2005,:Hla2004,
:Ian2006,:Nie2002,:Sim2007,:May2004,:Duj2002,:Mar2006,:Kno2002,:Wah2005,:Kra2006}.
Of course, each of these techniques has its specific advantages
and drawbacks. For instance, due to their rigid configuration,
MCBJs are very stable from a mechanical point of view;
consequently, experiments with variable temperatures can be
performed.  However, MCBJs have no lateral scanning ability and it
is very difficult to get information on the detailed
microscopic structure of the contacts (see above).
Electromigration techniques allow for building three-terminal
junctions in a straightforward way, but again the interface to the
molecules is not very well defined.
\tmfloatsmallfirst{\resizebox{\columnwidth}{!}{\includegraphics[
width=0.1\columnwidth]{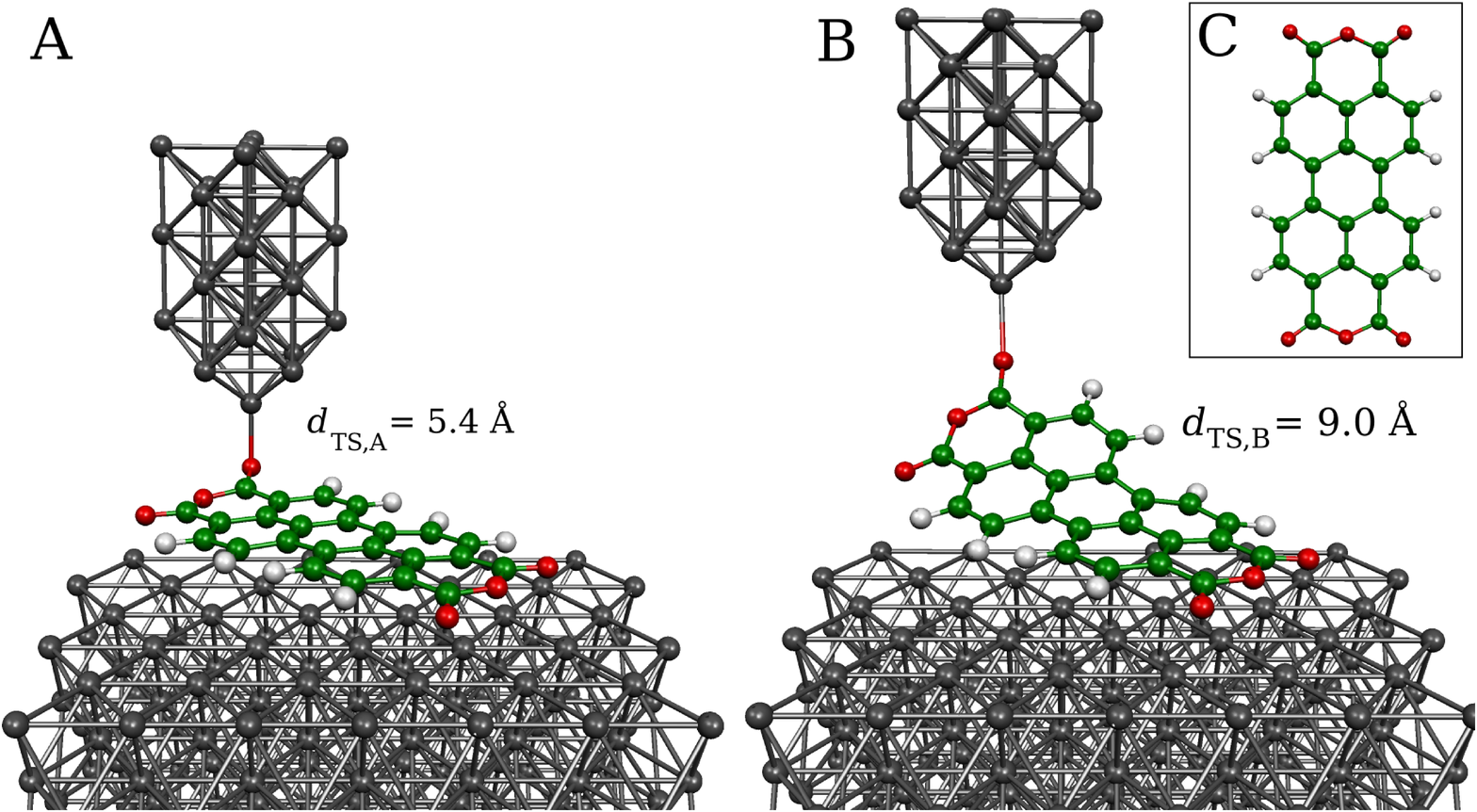}}}{\label{fig:lifting}
Relaxed geometries for two different tip substrate distances $d_{\mathrm{TS}}$
(A and B); top view of a PTCDA molecule (C).}

In contrast to the above techniques, STM-based transport
experiments in the tip-molecule-surface geometry -- by virtue of
the STM imaging capability -- allow for experimental access to the
atomic junction structure. However, STM junctions are mechanically
less stable and, if operated in the tunnelling regime, they yield
intrinsically asymmetric molecular junctions. Moreover, in an STM
junction electrical gating is often not available, because it is
very difficult to realize a third electrode in the vicinity of tip
and sample. Finally, temperature dependent transport experiments
are more difficult because of thermal drifts of the
STM tip with respect to the sample surface. These problems
notwithstanding, its is believed that STM-based transport
experiments are a worthwhile alternative to MCBJ and EM
experiments, because they offer unique advantages, most
importantly the structural control. As
such, STM-based transport experiments are complementary to the
other two techniques. Indeed, the STM tip cannot only address
individual molecules, but also specific sub-molecular groups or
even individual
atoms~\cite{:Nee2007,:Str2004,:Eig91,:Eig90,:Lee99,:Lim2005,:Rep2005a}.
Moreover, it must be noted here that some of the disadvantages of
the STM approach can in fact be overcome by using the STM in a
slightly unconventional fashion: firstly, the contact asymmetry
can be surmounted by contacting the molecule with the tip;
this contact may either lead to the formation of a strong
covalent chemical bond which survives tip retraction (``chemical
contact")~\cite{:Tem2008} or be purely ``mechanical" in
the sense that on tip retraction the contact between tip and
molecule is opened again~\cite{:Nee2007}. Secondly, in the case
of chemical contacts the tip can also be used for mechanical (in
contrast to capacitive) gating by changing the effective coupling
strengths between molecule and leads through variations of the
vertical tip position~\cite{:Tem2008}. This might
potentially allow one to scan different physical regimes ranging from
weak coupling (Coulomb blockade), through intermediate coupling
(Kondo physics) up to strong coupling (coherent transport).
By its capability of manipulation, STM thus allows for tuning
of the transport characteristics of individual molecules.

As a way out of the dilemma concerning reproducible molecular
electronics experiments mentioned above, we use here a two-step
approach based on commensurate, highly ordered molecular layers on
metal surfaces \cite{:Tem2008, :Sou2008}. In step 1, the powerful
armoury of surface science is employed to characterize the
structural and electronic properties of the metal-molecule
contact. Once step 1 has been completed, the tip of a
low-temperature STM is used to establish a covalent contact with
an individual molecule of the highly ordered monolayer (step 2).
Because of the excellent imaging properties of the STM, it is in
fact possible to select the part of the molecule which is
contacted with very high accuracy. In this way a structurally
well-defined molecular wire is realized which can be characterized
with respect to its transport properties. In favorable cases, it
may even be possible to apply a mechanical gating of the wire by
tip movements \cite{:Tem2008}.

This paper is organized as follows: Sects.~\ref{sect:STM} presents
the experimental strategy and the most important results of a proof-of-principle
experiment; Sect.~\ref{sect:JuncConf}-\ref{sect:MethSys} illustrate
how the interface from these experiments to \textit{ab initio}
theory can be devised.

\section{STM experiments}\label{sect:STM}
We have contacted a PTCDA
(4,9,10-perylene-tetracarboxylic-dianhydride) molecule with an STM
tip, peeling it off the Ag(111) surface and measuring its
transport properties in the process. PTCDA/Ag(111) is a
prototypical model interface, for which the following issues have
been addressed by previous work \cite{:Tau2007}:
\tmfloatsmalltop{\resizebox{\columnwidth}{!}{\includegraphics[
width=0.1\columnwidth]{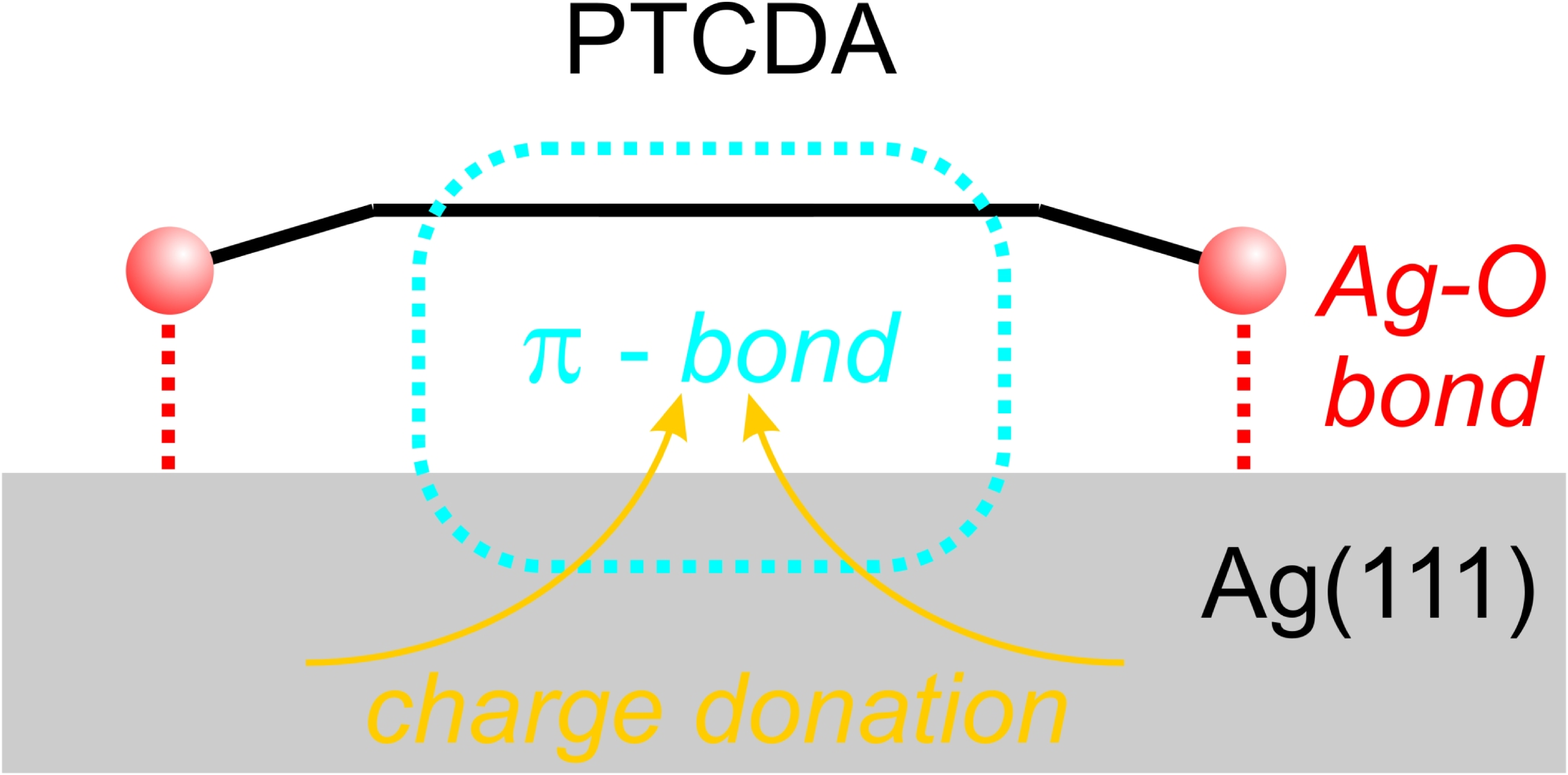}}}{\label{fig:Bond}
Schematic representation of the bonding interaction of PTCDA with
Ag(111) \cite{:New69, :Roh2007}.}(1) The precise geometric and
electronic structure of the adsorbed monolayer, including
site-specific effects and influence of the intermolecular
interactions, have been determined (cf.~the bonding model in
Fig.~\ref{fig:Bond}). (2) The mechanisms of the chemical bonding
of PTCDA to Ag(111), including orbital hybridization, charge
transfer, local versus extended bonds, and internal molecular
distortion have been analyzed. (3) The image formation in the STM
has been simulated from first principles. (4) Various density
functionals, including GGA and LDA, have been evaluated for the
PTCDA/Ag(111) adsorption system. In particular, it must be
noted here that according to the bonding model of
Fig.~\ref{fig:Bond} the carboxylic oxygen atoms are involved in
local Ag-O bonds. By their bond length they effectively define the vertical
distance of the perylene core above the metal surface~\cite{
:Kil2008}. This in turn controls the degree of metal-molecule
hybridization and charge transfer, because both proceed via the
extended $\pi$-electron system of the perylene backbone. At the
same time, it has been shown that the Ag-O bonds have only a
secondary influence on the electronic properties of the adsorbed
molecule~\cite{:Tem2008}. To first order, we can thus conclude that the Ag-O bonds act as (by themselves electronically
inactive) mechanical clamps which control the electron interaction
via the --as we will see below mechanically weaker --
perylene-metal bond.

PTCDA/Ag(111) is a system suited well to test the two-step approach
outlined at the end of the introduction, because of various reasions. Firstly, the bond
of PTCDA to the Ag(111) substrate is very well characterized,
i.e.~a lot of experiments with relevance to step 1 have already
been performed. Secondly, because of the coexistence of an
\emph{electronically active but mechanically weak extended bond}
and an \emph{electronically inactive but mechanically strong local
bond} (see Fig.~\ref{fig:Bond}), the molecule offers the
possibility for docking the STM tip to the mechanically active
terminal of the molecule and gradually cleaving the mechanically
weaker, electronic bond to the substrate by tip retraction from
the surface, thereby tuning the electronic coupling strength of
the molecule to the substrate. We shall see that in effect this
indeed amounts to a mechanical gating of transport through the
molecular wire. In this simple way a single molecule transport
experiment which combines a very high degree of structural control
at the contacts with electronic tunability of one of the contacts
is devised.
\tmfloatsmall{\resizebox{\columnwidth}{!}{\includegraphics{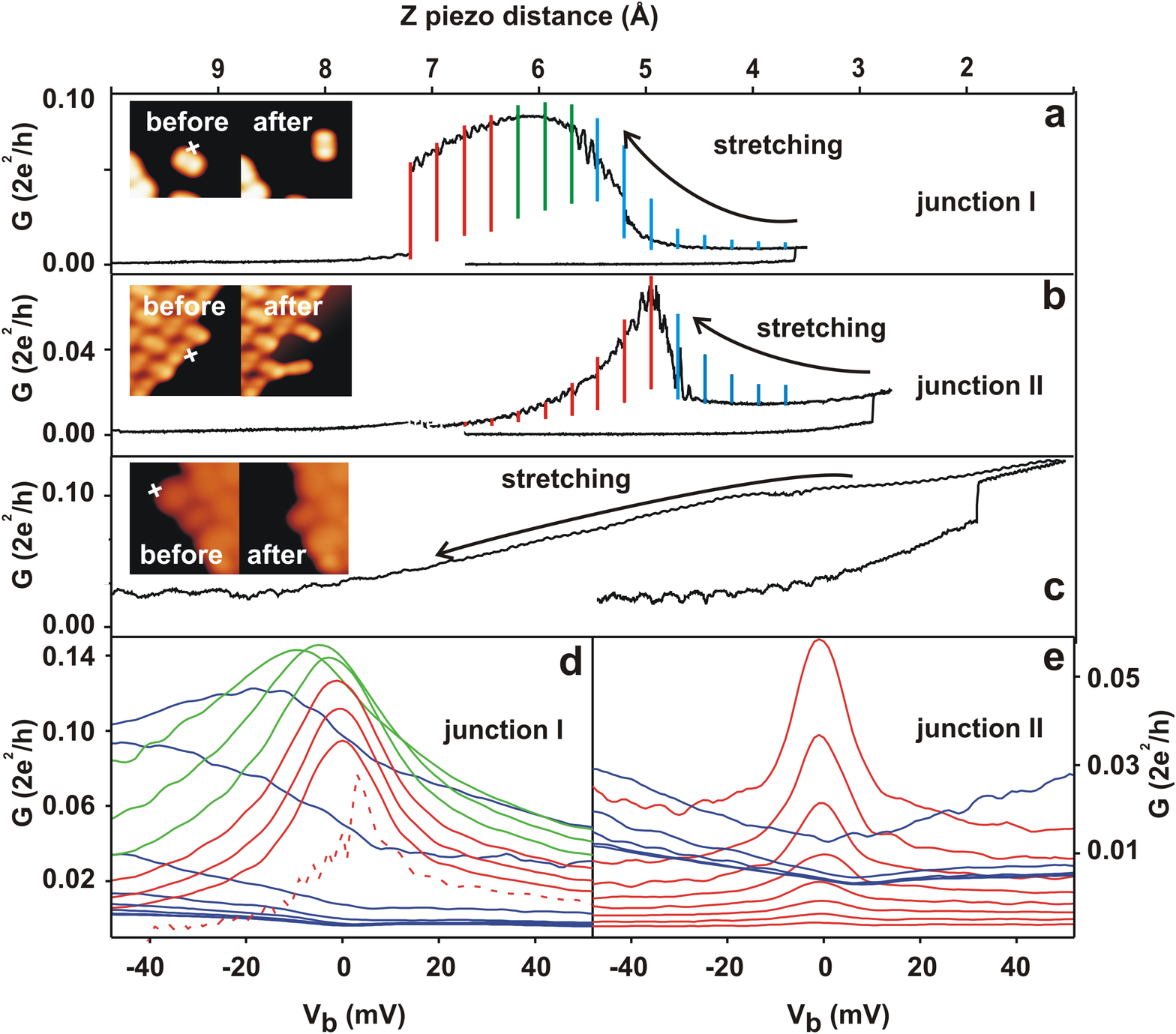}}}{\label{fig:Spectra}
Approach and retraction spectra of the conductance of the
stretched tip-PTCDA-Ag(111) (a, b) or tip-PTCDA-Au(111)(c)
transport junctions. Bias voltage $V_{b}$ = 2 mV for (a, b) and 4
mV for (c). Images in (a-c) show the investigated molecule before
and after the experiment, the cross indicates the point of
approach. The zero of the $z$ scale in (a-c) was established
according to the procedure described in Ref.~\cite{:Tem2008}.
Vertical bars in (a, b): range of differential conductance values from spectra in (d, e)). (d) Series of
$G=dI/dV$ vs. $V_{\mathrm{b}})$ spectra (smoothed) measured at $z$
positions indicated in (a) by vertical bars ($V_{\mathrm{mod}}$ =
4 mV, $\nu_{\mathrm{mod}}$ = 723 Hz). (e) Series of $G=dI/dV$ vs.
$V_{\mathrm{b}}$ spectra (smoothed) measured at $z$ positions
indicated in (b) by vertical bars ($V_{\mathrm{mod}}$ = 4 mV,
$\nu_{\mathrm{mod}}$ = 723 Hz). Colours in (d, e) code the regimes
in (a, b).}

First it must be established that it is indeed possible to contact
the PTCDA molecule at one of the carboxylic oxygens. This can be
concluded from the $z$-approach spectra, in which the conductance is
measured as a function of tip distance from the molecule
\cite{:Tem2008}. If the tip is approached above one of the
carboxylic oxygen atoms, a sudden increase of conductance by more
than one order of magnitude is observed at a tip distance of
approximately $2\,\mathrm{\AA}$ above the oxygen atom. This step
in the conductance is beyond the $z$-resolution of the STM. Since it
is only observed above the carboxylic oxygen atoms, we can
conclude that at the conductance step the oxygen atom flips up
into contact with the tip, to a new potential minimum. Further
analysis suggests that the tip-oxygen interaction can only be a
covalent bond \cite{:Tem2008}. For example, the reaction of the
tip is independent of the bias direction, which excludes an
electrostatic effect. Also, the tip-oxygen contact is strong
enough to remove the molecules from the surface. Finally, after
jumping into contact, the conductance reaches values which are
typical of covalent bonds (in the range of
$0.1\,G_{0}$)\cite{:Li2006} With these conductance values,
currents through a single molecule can become as large as a few
micro-amperes before the molecule is destroyed.

Once the tip-oxygen contact has been formed, it is possible to
retract the tip. We have mentioned already that in this way it is
possible to remove the molecule from the surface completely.
Analyzing the retraction more closely, we find that the
conductance remains larger than the corresponding tunnelling values
for retraction distances up to approximately $12\,\mathrm{\AA}$
\cite{:Tem2008}. Hence we can conclude that in this range of tip
positions the molecule is still present in the junction. Since the
length of the molecule is $12\,\mathrm{\AA}$, this indicates that
the molecule is really lifted from a parallel to a vertical
orientation by the tip, i.e.~the junction can indeed be stretched
out into a vertical wire. On tip retraction, two clearly
distinguishable regimes are observed: During the first
$4\,\mathrm{\AA}$, the retraction spectrum is smooth, but in the
range from $4\,\mathrm{\AA}$ to $12\,\mathrm{\AA}$ it becomes
rather jerky. It is clear that if the molecule is lifted up into
the vertical, its lower end must at some point slide over the
surface, and this is not expected to be a smooth but rather a
discontinuous movement.

The possibility to achieve a vertical orientation of the molecule
indicates that we are indeed able to modify the contact to the
substrate in wide bounds. Most notably, for PTCDA on Ag(111) we
invariably observe that the conductance goes through a maximum
within the first $4\,\mathrm{\AA}$ of tip retraction. To
understand the origin of this maximum, we have performed
experiments in which tip retraction has been halted at regular
intervals to measure full differential conductance
spectra in the range from $-50\,$mV to $50\,$mV. An example of
such a series of spectra is shown in Fig.~\ref{fig:Spectra}a, b,
d, e. As the tip is retracted, we observe in the spectra the
approach and sharpening of density of states from the left toward
the Fermi level. (We note here that for not too strongly
coupled wires, the energy-resolved transmission matrix and
--accordingly -- the differential conductance are proportional to
the electronic density of states in the wire). This density of
states must be assigned to the former LUMO of free PTCDA, which
serves as the bonding orbital responsible for the electronic
interaction (cf. the bonding sketch in Fig.~\ref{fig:Bond}). Its
behavior on junction stretching can be explained by
\emph{reversed chemisorption}, i.e.~dehybridization of the LUMO
from metal states (which leads to sharpening) and an upward shift
which follows the one-electron image potential of the occupied
orbital (negative charge!) above the metal surface
\cite{ :Tem2008}. Incidentally, this interpretation is confirmed by
analogous experiments with PTCDA on Au(111), which instead of the
conductance maximum always show a monotonous decrease of junction
conductance on tip retraction, as shown in Fig.~\ref{fig:Spectra}c. Since the interaction of PTCDA with Au(111) is merely physisorptive \cite{:Hen2007, :Ere2004}, little
or no charge transfer between molecule and the Au(111) substrate
occurs, hence the LUMO of PTCDA is never pulled below the Fermi
level. Lifting the molecule from the Au(111) surface thus can only
result in a further shift of the LUMO away from the Fermi level
(due to decreased polarization screening). The absence of the
maximum in the conductance curve for Au(111) thus indicates that
on Ag(111) it must be linked to the electronic interaction of the
molecule with the substrate, as described above.

At the end of the tip retraction experiment, the density of states
corresponding to the LUMO becomes pinned at the Fermi level
\cite{:Tem2008}. This cannot be explained in the framework of the
reversed chemisorption model, for which one would naively expect
the level to be pulled through the Fermi level, following the
one-electron image potential further. At least it is difficult to
rationalize in a one-particle model why the level should be pinned
exactly at the Fermi level, which corresponds to exactly one
electron in the orbital. However, if many-body physics is taken
into account, there is a mechanism that would naturally explain
the observed pinning. This is the Kondo effect. In
Ref.~\cite{:Tem2008} we have analyzed the possibility to explain
the pinning by the Kondo effect. To this end, an exact solution of
the Anderson Hamiltonian \cite{:And61}, using the numerical
renormalization group (NRG) \cite{:Kri80, :Pet2006, :Wei2007}, has
been used. Concerning the choice of the Anderson
Hamiltonian as a model for the present system, we recall that in
the formulation by Newns \cite{:New69} this Hamiltonian is widely
used as a model for molecular chemisorption on metal surfaces.
Indeed it is known from our earlier work that for
PTCDA/Ag(111)-chemisorption the LUMO of PTCDA behaves essentially
as a local level in the Newns-Anderson model \cite{:Kra2006a,
:Sou2008, :Tau2007}. Accordingly, the Anderson Hamiltonian is an
appropriate model for PTCDA/Ag(111), if the role of many-particle
correlations is to be analyzed.

The NRG calculation shows that pinning of the bonding orbital at
the Fermi level may indeed be observed even outside the fully
developed Kondo regime, i.e. at the border to the mixed
valence regime~\cite{:Tem2008}. This is an important finding,
because our data suggest that on tip retraction the
tip/PTCDA/Ag(111) wire remains on the borderline between
the Kondo and mixed valence regimes. This can be concluded from
the fact that we do not observe a very strong sharpening of the
zero bias feature with respect to the single particle peak. (The
observed sharpening amounts to a factor of approximately 2 only.) As a
consequence, we also do not resolve the two single particle peaks
at $\epsilon_0$ (singly occupied LUMO) and $\epsilon_0+U$ (doubly
occupied LUMO) separately, but rather a composite peak which
consists of the moderately sharpened many body peak in the center
and the two single particle peaks in the left and right tails. In
this context it is the most significant finding of the NRG
solution that a pinning of the composite resonance to the Fermi
level is indeed observed in the borderline regime between mixed
valence and Kondo physics -- in a parameter range that is
suggested by our experiments. The experiment in
Fig.~\ref{fig:Spectra} shows that before entering the pinning
regime $\epsilon_0$ roughly amounts to $40$\,meV. Note that the
order of magnitude of the intra-orbital Coulomb repulsion which we
have employed ($100\,$meV) has been corroborated by a many-body
calculation using the GW formalism, which yields $U=0.25\,$eV for
PTCDA on the silver substrate. This value includes a large
contribution of the screening by the metal surface: for the free
molecule a $U$ of $3.5\,$eV has been calculated.

In conclusion, the Kondo scenario in our experiments is supported
by strong circumstantial evidence. It must be stressed, however,
that an experimental test of great significance cannot be performed, since we
cannot stabilize the stretched junction at elevated temperatures
for times long enough to measure the conductance peak at zero
bias. This means that the expected decay of the Kondo peak at
temperatures around the Kondo temperature (approximately $120\,$K in our
case) cannot be verified \cite{:Tem2008}. However, we note here that in an
experiment with C$_{60}$ in a mechanically controlled break
junction \cite{:Par2006}, in which the Au contacts transfer charge into the
molecule and make the LUMO half-filled, a very similar zero bias
conductance peak has been observed on opening the junction, and in
that case it could be verified that the peak indeed decays
following the dependence predicted for a Kondo peak. Because the
transport physics in the experiment by Parks \textit{et al.}~is
very similar to the experiment on PTCDA/Ag(111) reported here, it
seems permissible to invoke an analogous temperature behavior for
our junctions, too. Incidentally, in the experiments by Parks
\textit{et al.}~only the pinning regime was observed. The
cross-over from the single-particle to the many-particle regime
was not reported there.

In spite of the internal consistency of the Kondo scenario
outlined above, a final \emph{proof }is needed. Because of its
structurally well-defined nature, the present system offers the
unique possibility to provide this proof by a comparison between
theory and experiment: Because experimentally the structure of the
PTCDA/Ag(111) adsorbate-substrate complex is well known, and
because its structural and electronic properties have reliably
been calculated by DFT, we can perform transport calculations of
the realistic junction. If this is done at various levels of
sophistication, the physics of the pinning can be illuminated: if
a \emph{single-particle} calculation does not show pinning, it is
very likely that the pinning is a \emph{many-body} phenomenon.
Conversely, if pinning is already observed at the single-particle
level, the assumption of Kondo physics may not be necessary,
although the mechanism outlined above should still remain valid;
its contribution to the observable behavior could then just be
dominated by some other mechanism.

In order to realize this program, two types of calculations are
necessary: First, starting from the experimentally verified
structure of the relaxed junction (no tip retraction), the
structural and electronic evolution of the junction on tip
retraction must be calculated. Secondly, on the basis of the
equilibrium structures of the stretched junction, transport
calculations may be performed. In the next section we shall turn to
a DFT simulation of the process of lifting the molecule off the
surface with the retracting STM tip. The results of this
simulation will fully corroborate the qualitative picture with
respect to contact formation and peel-off deduced from the
experiments.

\section{Junction atomistic configuration}\label{sect:JuncConf}
The fundamental questions to be answered by the DFT simulation of
the junction stretching process are the following. Firstly: can we verify the
flipping up of the carboxylic oxygen atom toward the tip and the
formation of a chemical bond between the two? Secondly: what
happens to the molecule structurally during tip retraction?

The geometrical configuration of the present system, consisting of
a metal substrate, molecule, and STM tip, can successfully be
described using density-functional theory (DFT). DFT, as realized
by the SIESTA package \cite{SIESTA,SIESTA1,SIESTA2}, has already
been used by some of the present authors to investigate the adsorption of a
monolayer of PTCDA on the Ag(111) surface
\cite{:Hau2005,:Kra2006,:Roh2007}. It has turned out that PTCDA
adsorbs on bridge sites above the substrate, at an equilibrium
height of about 2.8 \AA, characterized by a nearly flat geometry
with a slight molecular distortion. All these features are confirmed
by the corresponding experimental data \cite{:Hau2005,:Kra2006},
including the partial occupation of the molecular LUMO state by
electron charge transfer from the substrate. One should note that
DFT as such does not appear to be fully appropriate for the
adsorption of PTCDA on Ag(111) which apart from the chemical
interaction has a sizable contribution from physisorption, because
long-range van der Waals correlation effects are absent from in
DFT~\cite{:Roh2007}. Nonetheless, the theoretical results
obtained in Refs. \cite{:Hau2005,:Kra2006,:Roh2007} are in close
agreement with the measured data, thus justifying the use of DFT
for the questions to be studied here. This holds in particular for
the local-density approximation (LDA) to the exchange-correlation
functional, which yields much better results than the generalized
gradient approximation (GGA), which fails to bind the molecule to
the surface.

Based on the good results obtained previously for PTCDA on Ag(111)
\cite{:Roh2007}, we here proceed to include an STM tip into the
DFT calculation. Since we are not only interested in STM images
(which can be described successfully in terms of a Tersoff-Hamann
approach, i.e. without explicit inclusion of the tip
\cite{:Roh2007}), we focus on the mechanical interaction at close
distances, i.e. we allow for the relaxation of the adsorbed
molecules under the influence of forces exerted by the tip and
vice versa. To this end, the tip is included as a pyramidal
structure of ten tungsten atoms. In close relation to the
experimental setup described in Ref.~\cite{:Tem2008}, we place the
tip above the corner of the molecule, with the tip apex atom
exactly above one of the carboxylic oxygen atoms. A chemical bond
between the carboxylic oxygen atom and the apex tungsten atom
forms, with a bond length of 2.1 \AA. The calculations thus fully
confirm the conclusion drawn previously from our experimental
data. Thereafter, the STM tip is lifted up in steps of 0.2 \AA.
After each movement of the tip, the molecule is allowed to relax
and follow the tip. One should note that due to the large
complexity of the molecule it requires up to 100 geometric
relaxation steps until the molecule has adjusted to the retreated
tip. Fig.~\ref{fig:lifting} shows two typical relaxed structures
of this procedure. In configuration A, the molecule is still close
to its free adsorption structure. The carboxylic oxygen atom at
the left-hand side of the molecule, however, has formed the
chemical bond to the tungsten tip, which has a position of 5.4 \AA
\, above the Ag surface, corresponding to the initial phase of tip retraction. Upon retreating the tip further from the
surface, the carboxylic oxygen atom remains chemically bonded to
the tip and follows it, causing the molecule to bend and peel off
the surface. Fig.~\ref{fig:lifting}~B displays the configuration
after retreating the tip by 3.6 \AA.

The simulation reveals that the junction structure is
controlled by three competing energies and interactions. (i) The
bond between the carboxylic oxygen and the tip apex atom appears
to be quite stiff and to be of a covalent nature. (ii) The molecule is
still bonded to the substrate by a delocalized interaction
mechanism (cf. Fig.~\ref{fig:Bond}). This mechanism is driven by
the electron charge transfer from the substrate into the molecular
LUMO state, which is delocalized over the entire perylene core of
the molecule \cite{:Roh2007}. This causes attraction between every
part of the molecule and the substrate. (iii) The perylene core is
relatively soft and can easily be deformed. The sum of all three
interaction mechanisms finally yields the curved structure of the
molecule in Fig.~\ref{fig:lifting}, with its right-hand part still arranged flat on the
substrate and the left-hand part lifted off. Apparently, the
attraction of the right-hand part of the molecule to the surface
is still stronger than the energy cost of bending the molecule.
Again this is in good agreement with the behavior deduced from
experiments. At present, the calculations have not yet reached the regime in
which the molecule slides across the surface.

\tmfloatsmall{\resizebox{\columnwidth}{!}{\includegraphics[
width=0.1\columnwidth]{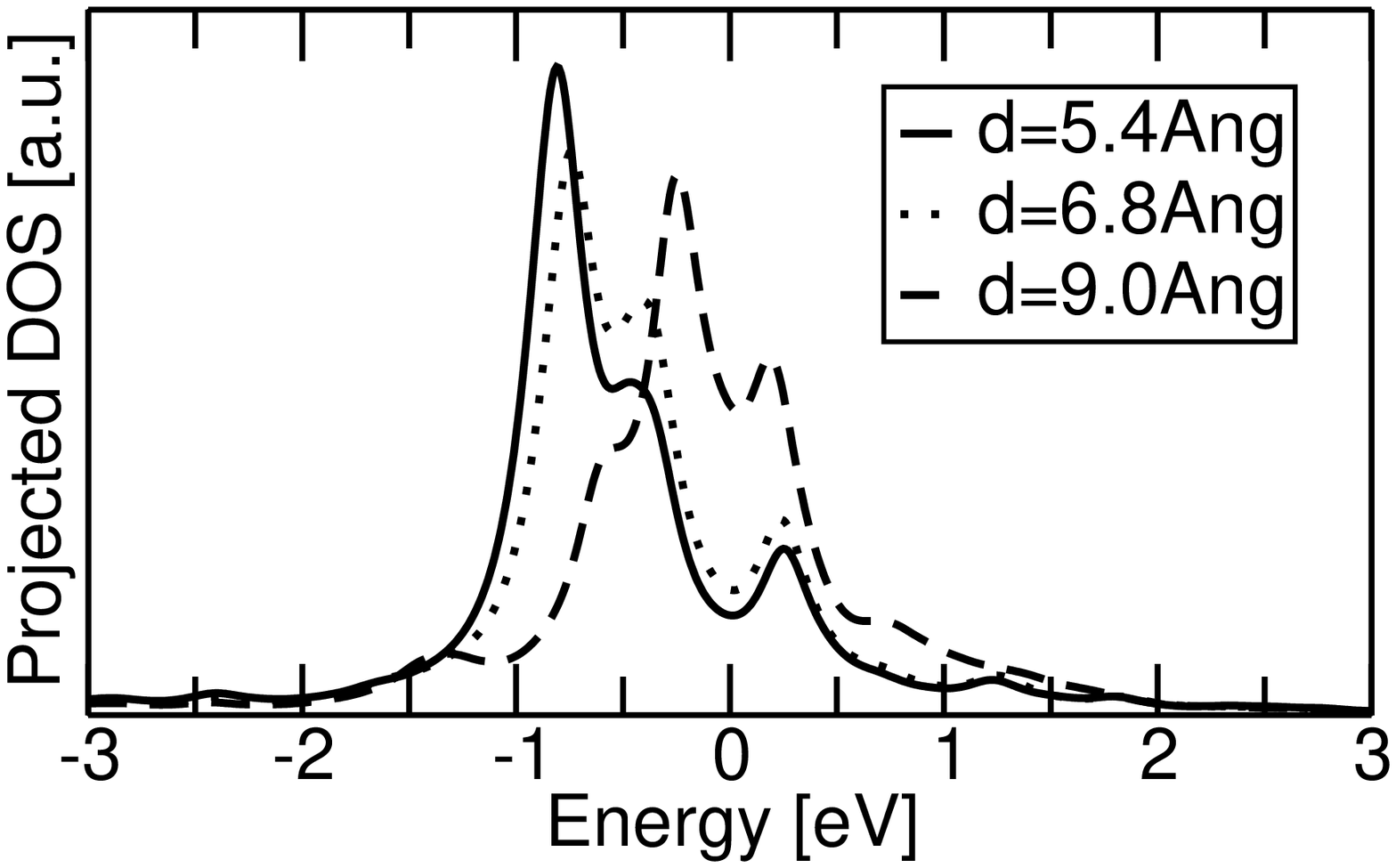}}}{\label{fig:pdos}
Density of states ofthe LUMO state, projected onto the electronic 
states of the substrate-adsorbate complex, for configurations obtained 
at tip-surface distances of 5.4 \AA, 6.8 \AA, and 9.0 \AA.}
One important feature during the peeling process concerns the 
occupation of the LUMO state. This can be monitored by 
projecting the gas-phase LUMO wave function onto the states of the
substrate-adsorbate system. 
Results for three configurations (at tip-surface distances of 5.4 \AA,
6.8 \AA, and 9.0 \AA) are displayed in Fig. \ref{fig:pdos}.
These are the configurations for which transport calculations 
will be discussed further below.
It is clearly visible that, in accordance with the mechanism as 
described above, the LUMO is shifted to higher energy during the 
peeling process, and at 9.0 \AA \, tip height it has already a substantial 
density of states above the Fermi level.
This is accompanied by successive de-population and spectral 
sharpening. The completely removed molecule would have a sharp 
LUMO above the Fermi level of the substrate.

The geometries as determined by the DFT simulation of the lift-off are the basis for the transport calculations described in the next section.

\section{DFT-based coherent transport calculations}\label{sect:MethSys}
As a further step towards a deeper understanding of the transport properties
described in Sec.~\ref{sect:STM}, we perform transport simulations through a
junction composed of substrate, molecule and STM tip. Even though model-based approaches often give valuable insight into
underlying transport mechanisms \cite{ :Guti2005, :Pump2007, :Song2007, :Ryn2008}, a realistic, material-specific description can
only be achieved by an atomistic study. Given the size of the system, an
approximate density functional parametrized tight-binding method, DFTB, is used
to account for the electronic structure of the hybrid system. The input
geometries for these calculations are the fully relaxed ones obtained using the
DFT package SIESTA as described above in Sec.~\ref{sect:JuncConf}, to obtain optimal reliability of the structure. 

As a first step in the transport simulations, we show in the present study as a characterizing
feature the transmission function $T \left( E \right)$ at zero bias voltage
for different distances between the substrate and the STM tip. In
this way one can follow the evolution of the molecular levels (peaks in
the transmission spectra) and the electronic coupling to the
contacts (metallic surface and STM tip, respectively) which is
reflected by the width of these peaks. To do so, we use as
computational method the quantum transport implementation
gDFTB~\cite{gDFTB1, gDFTB2}, which is a combination of the
non-equilibrium Green function technique to describe the transport
process and a DFT-based two-center charge
self-consistent tight-binding method, DFTB~\cite{DFTB1}, to
account for the electronic structure of the extended molecular
system (see Fig.~\ref{fig:lifting}). During the transport
calculation, starting from the system Hamiltonian, a charge self-consistency loop over the non-equilibrium Green function, the
density matrix and the real space charge density is performed
under the non-equilibrium conditions imposed by the applied bias
voltage and the attached contacts. The resulting electric field is
calculated via solving the Poisson equation at any loop cycle.
The electric contacts are modeled as ideal semi-infinite wires,
which are projected by decimation techniques onto an effective
surface Green function. As the transport property considered here,
the coherent transmission function is given by
\begin{equation}
T\left( E\ \right)= \mathrm{Tr} \left[ \Gamma_{\mathrm{1}} G^{\mathrm{r}}
\Gamma_{\mathrm{2}} G^{\mathrm{a}}\right],
\end{equation}
where $\Gamma_{\mathrm{1}}$ and $\Gamma_{\mathrm{2}}$ describe the
coupling between the electrodes and the extended molecule at the
contact layers.  $G^{\mathrm{r}}$ and  $G^{\mathrm{a}}$ are the
retarded and advanced Green function of the extended molecule
which are calculated as the result of the self-consistency loop.

It has to be stressed that the presented calculations are
single-particle ones and cannot capture typical Kondo physics
features or charging effects. However, the results give deep
insight into the electronic structure and its evolution during the lifting
process and provide hints as to whether the
observed experimental features can be explained in a
single-particle picture or need a more sophisticated explanation.

\tmfloatsmall{\resizebox{\columnwidth}{!}{\includegraphics{2008_01_15_APA_PuTaRoCu_Fig5.eps}}}{\label{fig:Trans}Transmission at zero bias
as a function of energy for different tip substrate distances
$d_{\mathrm{TS}}$.}
In the transport calculations, a single PTCDA molecule, one surface layer
consisting of 38 atoms and 10 atoms of the top of the STM tip are
taken as the extended molecule to be treated fully within DFTB for
taking into account charge transfer processes between the surface
and the molecule at all stages of the lifting process. Both leads are
modeled by semi-infinite silver wires. It has to be noted that finite size
effects may lead to a quantitative  difference compared to studies of the system
taking into account periodic boundary conditions which are
currently carried out. The parameter sets used in this study were the ones
available from the DFTB project~\cite{DFTB1, :Szu2003, :Szu2004, DFTB}.
Fig.~\ref{fig:Trans} shows the zero bias transmission function $T$
as a function of the incoming particle energy $E$ for different
tip-substrate distances. It can clearly be seen that the spectrum,
especially the peak closest to the Fermi energy, moves in a
reversed chemisorption process in a positive energy direction as
observed in experiment. For the largest distance investigated in the present study,
{\textit{i.e.}} 9.0~\AA, the peak is located closely below the
Fermi energy. The further evolution of the position of this peak
with larger distances (the molecule lifted further from the
surface) is of great interest as it could show whether the level
crosses the Fermi energy or if it remains pinned there. The
relaxation of the system for this parameter regime is currently
under investigation. Furthermore it is evident from
Fig.~\ref{fig:Trans} that the width of the peaks in the spectra is
decreased for larger distances, as, due to a reduction of the
overlap of the molecular $\pi$-system with the metal electronic
system, the electronic coupling strength and hence the injection
rate between the molecule and the metallic substrate is reduced
(Breit-Wigner physics).

\section{Conclusions and Outlook}\label{sect:Conclusios}

Using quantum transport methods, we have shown the motion of the
energy levels in positive direction, especially of the frontier
orbitals when lifting the molecule from the substrate.

In continuation of the present work, it is necessary to explore
stable system setups with larger tip-substrate distances to follow
the whole trace of the molecule during the lifting process as
performed in the experiments, where also lateral movement of the
molecules on the surface is observed. Furthermore, an
investigation of the different behavior and properties comparing
isolated molecules on a surface to a monolayer of PTCDA is
presently performed. To finally clarify the origin of the
experimentally observed features, it is necessary to find clear
relations between the important magnitudes governing
transport and the experimental input parameter, the tip substrate
distance. In detail, these are the charging energy $U$, the
coupling strengths between molecule and substrate, described
by $\Gamma$, and the level energy $\epsilon$. Depending on their
functional dependence during the lifting process, it will be
possible to decide whether a single particle picture is sufficient
to describe the observations or if a Kondo picture needs to be
applied.

Quite apart form the question whether the Kondo scenario is
applicable for our model system, our proof-of-principle experiment
and the simulation program, carried out here for PTCDA, provide a
benchmark for the comparison of a single-molecule transport
experiment and \textit{ab-initio} transport simulations, because
hardly ever is structural information on the
transport junction available that is so detailed.

\section{Acknowledgements}
This work was partially funded by the Volkswagen Foundation
(contract I/78~340), the German Research Foundation (DFG) Priority
Program 1243 ``Quantum transport at the molecular scale''. The
Center for Information Services and High Performance Computing
(ZIH) at the Dresden University of Technology provided computation
time. We are indebted to Aldo Di Carlo and Alessandro Pecchia for
the gDFTB code and to Alessio Gagliardi, Rafael Guti\'errez and Bo
Song for fruitful discussions.

%\bibliographystyle{revtex}
%\bibliography{2008_01_15_APA_PuTaRoCu}

\begin{thebibliography}{10}
\providecommand*{\bibinfo}[2]{#2}
\providecommand*{\eprint}[1]{#1}
\providecommand*{\url}[1]{#1}
\bibitem{:Cun2005}
\bibinfo{author}{G.~Cuniberti, G.~Fagas} and \bibinfo{author}{K.~Richter},
  \bibinfo{title}{\emph{Introducing Molecular Electronics: A brief overview}},
  \bibinfo{volume}{vol. 680 of \emph{Lecture Notes in Physics}}
  (\bibinfo{publisher}{Springer}, \bibinfo{year}{2005}).
\bibitem{AR74a}
\bibinfo{author}{A.~Aviram} and \bibinfo{author}{M.~A. Ratner},
  \bibinfo{journal}{Chem.\ Phy.\ Lett.} \bibinfo{volume}{\textbf{29}}(2),
  \bibinfo{pages}{277} (\bibinfo{date}{1974}).
\bibitem{nitzan03}
\bibinfo{author}{A.~Nitzan} and \bibinfo{author}{M.~A. Ratner},
  \bibinfo{journal}{Science} \bibinfo{volume}{\textbf{300}},
  \bibinfo{pages}{1384} (\bibinfo{date}{2003}).
\bibitem{:Joa2005}
\bibinfo{author}{C.~Joachim} and \bibinfo{author}{M.~A. Ratner},
  \bibinfo{journal}{Proc.\ Natl.\ Acad.\ Sci.\ USA}
  \bibinfo{volume}{\textbf{102}}, \bibinfo{pages}{8800} (\bibinfo{date}{2005}).
\bibitem{:Val2007}
\bibinfo{author}{M.~{del Valle}}, \bibinfo{author}{R.~Guti{\'e}rrez},
  \bibinfo{author}{C.~Tejedor}, and \bibinfo{author}{G.~Cuniberti},
  \bibinfo{journal}{Nature Nanotech.} \bibinfo{volume}{\textbf{2}},
  \bibinfo{pages}{176} (\bibinfo{date}{2007}).
\bibitem{:Ke2005}
\bibinfo{author}{S.-H. Ke}, \bibinfo{author}{H.~U. Baranger}, and
  \bibinfo{author}{W.~Yang}, \bibinfo{journal}{J. Chem. Phys.}
  \bibinfo{volume}{\textbf{123}}, \bibinfo{pages}{114701}
  (\bibinfo{date}{2005}).
\bibitem{:Kau2008}
\bibinfo{author}{C.-C. Kaun} and \bibinfo{author}{T.~Seideman},
  \bibinfo{journal}{Phys. Rev. B} \bibinfo{volume}{\textbf{77}},
  \bibinfo{pages}{033414} (\bibinfo{date}{2008}).
\bibitem{:Tem2008}
\bibinfo{author}{R.~Temirov}, \bibinfo{author}{A.~C. Lassise},
  \bibinfo{author}{F.~Anders}, and \bibinfo{author}{F.~S. Tautz},
  \bibinfo{journal}{Nanotechnology} \bibinfo{volume}{\textbf{19}},
  \bibinfo{pages}{065401} (\bibinfo{date}{2008}).
\bibitem{:Nee2007}
\bibinfo{author}{N.~N\'eel}, \bibinfo{author}{J.~Kr\"oger},
  \bibinfo{author}{L.~Limot}, \bibinfo{author}{T.~Frederiksen},
  \bibinfo{author}{M.~Brandbyge}, and \bibinfo{author}{R.~Berndt},
  \bibinfo{journal}{Phys.\ Rev.\ Lett.} \bibinfo{volume}{\textbf{98}},
  \bibinfo{pages}{065502} (\bibinfo{date}{2007}).
\bibitem{:Nee2007a}
\bibinfo{author}{N.~N\'eel}, \bibinfo{author}{J.~Kr\"oger},
  \bibinfo{author}{L.~Limot}, \bibinfo{author}{J.~Palotas},
  \bibinfo{author}{W.~Hofer}, and \bibinfo{author}{R.~Berndt},
  \bibinfo{journal}{Phys.\ Rev.\ Lett.} \bibinfo{volume}{\textbf{98}},
  \bibinfo{pages}{016801} (\bibinfo{date}{2007}).
\bibitem{:Kro2007}
\bibinfo{author}{J.~Kr\"oger}, \bibinfo{author}{H.~Jensen}, and
  \bibinfo{author}{R.~Berndt}, \bibinfo{journal}{New J. Phys.}
  \bibinfo{volume}{\textbf{9}}, \bibinfo{pages}{153} (\bibinfo{date}{2007}).
\bibitem{:Lim2005}
\bibinfo{author}{L.~Limot}, \bibinfo{author}{J.~Kr\"oger},
  \bibinfo{author}{R.~Berndt}, \bibinfo{author}{A.~Garcia-Lekue}, and
  \bibinfo{author}{W.~A. Hofer}, \bibinfo{journal}{Phys.\ Rev.\ Lett.}
  \bibinfo{volume}{\textbf{94}}, \bibinfo{pages}{126102}
  (\bibinfo{date}{2005}).
\bibitem{:Yaz1996}
\bibinfo{author}{A.~Yazdani}, \bibinfo{author}{D.~M. Eigler}, and
  \bibinfo{author}{N.~D. Lang}, \bibinfo{journal}{Science}
  \bibinfo{volume}{\textbf{272}}, \bibinfo{pages}{1921} (\bibinfo{date}{1996}).
\bibitem{:Dre2005}
\bibinfo{author}{M.~Dreher}, \bibinfo{author}{F.~Pauly},
  \bibinfo{author}{J.~Heurich}, \bibinfo{author}{J.~C. Cuevas},
  \bibinfo{author}{E.~Scheer}, and \bibinfo{author}{P.~Nielaba},
  \bibinfo{journal}{Phys.\ Rev.\ B} \bibinfo{volume}{\textbf{72}},
  \bibinfo{pages}{075435} (\bibinfo{date}{2005}).
\bibitem{:Sch97}
\bibinfo{author}{E.~Scheer}, \bibinfo{author}{P.~Joyez},
  \bibinfo{author}{D.~Esteve}, \bibinfo{author}{C.~Urbina}, and
  \bibinfo{author}{M.~H. Devoret}, \bibinfo{journal}{Phys.\ Rev.\ Lett.}
  \bibinfo{volume}{\textbf{78}}, \bibinfo{pages}{3535} (\bibinfo{date}{1997}).
\bibitem{:Sch98}
\bibinfo{author}{E.~Scheer}, \bibinfo{author}{N.~Agrait},
  \bibinfo{author}{J.~C. Cuevas}, \bibinfo{author}{A.~L. Yeyati},
  \bibinfo{author}{B.~Ludoph}, \bibinfo{author}{A.~Martin-Rodero},
  \bibinfo{author}{G.~R. Bollinger}, \bibinfo{author}{J.~M. van Ruitenbeek},
  and \bibinfo{author}{C.~Urbina}, \bibinfo{journal}{Nature}
  \bibinfo{volume}{\textbf{394}}, \bibinfo{pages}{154} (\bibinfo{date}{1998}).
\bibitem{:Elb2005}
\bibinfo{author}{M.~Elbing}, \bibinfo{author}{R.~Ochs},
  \bibinfo{author}{M.~Koentopp}, \bibinfo{author}{M.~Fischer},
  \bibinfo{author}{C.~von Hanisch}, \bibinfo{author}{F.~Weigend},
  \bibinfo{author}{F.~Evers}, \bibinfo{author}{H.~B. Weber}, and
  \bibinfo{author}{M.~Mayor}, \bibinfo{journal}{Proc.\ Natl.\ Acad.\ Sci.\ USA}
  \bibinfo{volume}{\textbf{102}}(25), \bibinfo{pages}{8815}
  (\bibinfo{date}{2005}).
\bibitem{:Rei2002}
\bibinfo{author}{J.~Reichert}, \bibinfo{author}{R.~Ochs},
  \bibinfo{author}{D.~Beckmann}, \bibinfo{author}{H.~B. Weber},
  \bibinfo{author}{M.~Mayor}, and \bibinfo{author}{H.~v.~L\"ohneysen},
  \bibinfo{journal}{Phys.\ Rev.\ Lett.} \bibinfo{volume}{\textbf{88}},
  \bibinfo{pages}{176804} (\bibinfo{date}{2002}).
\bibitem{:Rei2003}
\bibinfo{author}{J.~Reichert}, \bibinfo{author}{H.~B. Weber},
  \bibinfo{author}{M.~Mayor}, and \bibinfo{author}{H.~v.~Lohneysen},
  \bibinfo{journal}{Appl.\ Phys.\ Lett.} \bibinfo{volume}{\textbf{82}},
  \bibinfo{pages}{4137} (\bibinfo{date}{2003}).
\bibitem{:Web2001}
\bibinfo{author}{H.~B. Weber}, \bibinfo{author}{R.~H\"aussler},
  \bibinfo{author}{H.~v.~L\"ohneysen}, and \bibinfo{author}{J.~Kroha},
  \bibinfo{journal}{Phys.\ Rev.\ B} \bibinfo{volume}{\textbf{63}},
  \bibinfo{pages}{165426} (\bibinfo{date}{2001}).
\bibitem{:Xie2004}
\bibinfo{author}{F.~Q. Xie}, \bibinfo{author}{L.~Nittler},
  \bibinfo{author}{C.~Obermair}, and \bibinfo{author}{T.~Schimmel},
  \bibinfo{journal}{Phys.\ Rev.\ Lett.} \bibinfo{volume}{\textbf{93}},
  \bibinfo{pages}{128303} (\bibinfo{date}{2004}).
\bibitem{:Smi2002}
\bibinfo{author}{R.~H.~M. Smit}, \bibinfo{author}{Y.~Noat},
  \bibinfo{author}{C.~Untiedt}, \bibinfo{author}{N.~D. Lang},
  \bibinfo{author}{M.~C. van Hemert}, and \bibinfo{author}{J.~M. van
  Ruitenbeek}, \bibinfo{journal}{Nature} \bibinfo{volume}{\textbf{419}},
  \bibinfo{pages}{906} (\bibinfo{date}{2002}).
\bibitem{:Agr2003}
\bibinfo{author}{A.~Agrait}, \bibinfo{author}{A.~Levy-Yeyati}, and
  \bibinfo{author}{J.~M. van Ruitenbeek}, \bibinfo{journal}{Phys. Rep.}
  \bibinfo{volume}{\textbf{377}}, \bibinfo{pages}{81} (\bibinfo{date}{2003}).
\bibitem{EM1}
\bibinfo{author}{H.~B. Heersche}, \bibinfo{author}{Z.~{de~Groot}},
  \bibinfo{author}{J.~A. Folk}, \bibinfo{author}{L.~P. Kouwenhoven},
  \bibinfo{author}{H.~S.~J. {van der Zant}}, \bibinfo{author}{A.~A. Houck},
  \bibinfo{author}{J.~Labaziewicz}, and \bibinfo{author}{I.~L. Chuang},
  \bibinfo{journal}{Phys.\ Rev.\ Lett.} \bibinfo{volume}{\textbf{96}},
  \bibinfo{pages}{017205} (\bibinfo{date}{2006}).
\bibitem{:Rep2005}
\bibinfo{author}{J.~Repp}, \bibinfo{author}{G.~Meyer},
  \bibinfo{author}{S.~Paavilainen}, \bibinfo{author}{F.~E. Olsson}, and
  \bibinfo{author}{M.~Persson}, \bibinfo{journal}{Phys.\ Rev.\ Lett.}
  \bibinfo{volume}{\textbf{95}}, \bibinfo{pages}{225503}
  (\bibinfo{date}{2005}).
\bibitem{:Sti98}
\bibinfo{author}{B.~C. Stipe}, \bibinfo{author}{M.~A. Rezaei}, and
  \bibinfo{author}{W.~Ho}, \bibinfo{journal}{Phys.\ Rev.\ Lett.}
  \bibinfo{volume}{\textbf{81}}, \bibinfo{pages}{1263 } (\bibinfo{date}{1998}).
\bibitem{:Sti99}
\bibinfo{author}{B.~C. Stipe}, \bibinfo{author}{M.~A. Rezaei}, and
  \bibinfo{author}{W.~Ho}, \bibinfo{journal}{Phys.\ Rev.\ Lett.}
  \bibinfo{volume}{\textbf{82}}, \bibinfo{pages}{1724 } (\bibinfo{date}{1999}).
\bibitem{:Cho2006}
\bibinfo{author}{B.-Y. Choi}, \bibinfo{author}{S.-J. Kahng},
  \bibinfo{author}{S.~Kim}, \bibinfo{author}{H.~Kim},
  \bibinfo{author}{H.~Won~Kim}, \bibinfo{author}{Y.~J. Song},
  \bibinfo{author}{J.~Ihm}, and \bibinfo{author}{Y.~Kuk},
  \bibinfo{journal}{Phys.\ Rev.\ Lett.} \bibinfo{volume}{\textbf{96}},
  \bibinfo{pages}{156106} (\bibinfo{date}{2006}).
\bibitem{:Yan2002}
\bibinfo{author}{H.~Yanagi}, \bibinfo{author}{K.~Ikuta},
  \bibinfo{author}{H.~Mukai}, and \bibinfo{author}{T.~Shibutani},
  \bibinfo{journal}{Nano Letters} \bibinfo{volume}{\textbf{2}},
  \bibinfo{pages}{951} (\bibinfo{date}{2002}).
\bibitem{:Bra2005}
\bibinfo{author}{K.-F. Braun} and \bibinfo{author}{S.-W. Hla},
  \bibinfo{journal}{Nano Letters} \bibinfo{volume}{\textbf{5}},
  \bibinfo{pages}{73 } (\bibinfo{date}{2005}).
\bibitem{:Hla2004}
\bibinfo{author}{S.-W. Hla}, \bibinfo{author}{K.-F. Braun},
  \bibinfo{author}{B.~Wassermann}, and \bibinfo{author}{K.-H. Rieder},
  \bibinfo{journal}{Phys.\ Rev.\ Lett.} \bibinfo{volume}{\textbf{93}},
  \bibinfo{pages}{208302} (\bibinfo{date}{2004}).
\bibitem{:Ian2006}
\bibinfo{author}{V.~Iancu}, \bibinfo{author}{A.~Deshpande}, and
  \bibinfo{author}{S.-W. Hla}, \bibinfo{journal}{Nano Lett.}
  \bibinfo{volume}{\textbf{6}}, \bibinfo{pages}{820} (\bibinfo{date}{2006}).
\bibitem{:Nie2002}
\bibinfo{author}{J.~Nieminen}, \bibinfo{author}{S.~Lahti},
  \bibinfo{author}{S.~Paavilainen}, and \bibinfo{author}{K.~Morgenstern},
  \bibinfo{journal}{Phys.\ Rev.\ B} \bibinfo{volume}{\textbf{66}},
  \bibinfo{pages}{165421 (2002)} (\bibinfo{date}{2002}).
\bibitem{:Sim2007}
\bibinfo{author}{V.~Simic-Milosevic}, \bibinfo{author}{M.~Mehlhorn},
  \bibinfo{author}{K.-H. Rieder}, \bibinfo{author}{J.~Meyer}, and
  \bibinfo{author}{K.~Morgenstern}, \bibinfo{journal}{Phys.\ Rev.\ Lett.}
  \bibinfo{volume}{\textbf{98}}, \bibinfo{pages}{116102}
  (\bibinfo{date}{2007}).
\bibitem{:May2004}
\bibinfo{author}{A.~J. Mayne}, \bibinfo{author}{M.~Lastapis},
  \bibinfo{author}{G.~Baffou}, \bibinfo{author}{L.~Soukiassian},
  \bibinfo{author}{G.~Comtet}, \bibinfo{author}{L.~Hellner}, and
  \bibinfo{author}{G.~Dujardin}, \bibinfo{journal}{Phys.\ Rev.\ B}
  \bibinfo{volume}{\textbf{69}}, \bibinfo{pages}{045409}
  (\bibinfo{date}{2004}).
\bibitem{:Duj2002}
\bibinfo{author}{G.~Dujardin}, \bibinfo{author}{A.~J. Mayne}, and
  \bibinfo{author}{F.~Rose}, \bibinfo{journal}{Phys.\ Rev.\ Lett.}
  \bibinfo{volume}{\textbf{89}}, \bibinfo{pages}{036802}
  (\bibinfo{date}{2002}).
\bibitem{:Mar2006}
\bibinfo{author}{M.~Martin}, \bibinfo{author}{M.~Lastapis},
  \bibinfo{author}{D.~Riedel}, \bibinfo{author}{G.~Dujardin},
  \bibinfo{author}{M.~Mamatkulov}, \bibinfo{author}{L.~Stauffer}, and
  \bibinfo{author}{P.~Sonnet}, \bibinfo{journal}{Phys.\ Rev.\ Lett.}
  \bibinfo{volume}{\textbf{97}}, \bibinfo{pages}{216103}
  (\bibinfo{date}{2006}).
\bibitem{:Kno2002}
\bibinfo{author}{N.~Knorr}, \bibinfo{author}{M.~A. Schneider},
  \bibinfo{author}{L.~Diekh\"oner}, \bibinfo{author}{P.~Wahl}, and
  \bibinfo{author}{K.~Kern}, \bibinfo{journal}{Phys.\ Rev.\ Lett.}
  \bibinfo{volume}{\textbf{88}}, \bibinfo{pages}{096804}
  (\bibinfo{date}{2002}).
\bibitem{:Wah2005}
\bibinfo{author}{P.~Wahl}, \bibinfo{author}{L.~Diekh\"oner},
  \bibinfo{author}{G.~Wittich}, \bibinfo{author}{L.~Vitali},
  \bibinfo{author}{M.~A. Schneider}, and \bibinfo{author}{K.~Kern},
  \bibinfo{journal}{Phys.\ Rev.\ Lett.} \bibinfo{volume}{\textbf{95}},
  \bibinfo{pages}{166601} (\bibinfo{date}{2005}).
\bibitem{:Kra2006}
\bibinfo{author}{A.~Kraft}, \bibinfo{author}{R.~Temirov},
  \bibinfo{author}{S.~K.~M. Henze}, \bibinfo{author}{S.~Soubatch},
  \bibinfo{author}{M.~Rohlfing}, and \bibinfo{author}{F.~S. Tautz},
  \bibinfo{journal}{Phys.\ Rev.\ B} \bibinfo{volume}{\textbf{74}},
  \bibinfo{pages}{041402(R)} (\bibinfo{date}{2006}).
\bibitem{:Str2004}
\bibinfo{author}{J.~A. Stroscio} and \bibinfo{author}{R.~J. Celotta},
  \bibinfo{journal}{Science} \bibinfo{volume}{\textbf{306}},
  \bibinfo{pages}{242} (\bibinfo{date}{2004}).
\bibitem{:Eig91}
\bibinfo{author}{D.~M. Eigler}, \bibinfo{author}{C.~P. Lutz}, and
  \bibinfo{author}{W.~E. Rudge}, \bibinfo{journal}{Nature}
  \bibinfo{volume}{\textbf{352}}, \bibinfo{pages}{600} (\bibinfo{date}{1991}).
\bibitem{:Eig90}
\bibinfo{author}{D.~M. Eigler} and \bibinfo{author}{E.~K. Schweizer},
  \bibinfo{journal}{Nature} \bibinfo{volume}{\textbf{344}},
  \bibinfo{pages}{524} (\bibinfo{date}{1990}).
\bibitem{:Lee99}
\bibinfo{author}{H.~J. Lee} and \bibinfo{author}{W.~Ho},
  \bibinfo{journal}{Science} \bibinfo{volume}{\textbf{286}},
  \bibinfo{pages}{1719} (\bibinfo{date}{1999}).
\bibitem{:Rep2005a}
\bibinfo{author}{J.~Repp}, \bibinfo{author}{G.~Meyer}, \bibinfo{author}{S.~M.
  Stojkovic}, \bibinfo{author}{A.~Gourdon}, and \bibinfo{author}{C.~Joachim},
  \bibinfo{journal}{Phys.\ Rev.\ Lett.} \bibinfo{volume}{\textbf{94}},
  (\bibinfo{date}{2005}).
\bibitem{:Sou2008}
\bibinfo{author}{S.~Soubatch}, \bibinfo{author}{R.~Temirov}, and
  \bibinfo{author}{F.~S. Tautz}, \bibinfo{journal}{Phys. Stat. Sol. A}
  \bibinfo{volume}{\textbf{205}}, \bibinfo{pages}{511} (\bibinfo{date}{2008}).
\bibitem{:Tau2007}
\bibinfo{author}{F.~S. Tautz}, \bibinfo{journal}{Prog. Surf. Sci.}
  \bibinfo{volume}{\textbf{82}}, \bibinfo{pages}{479} (\bibinfo{date}{2007}).
\bibitem{:New69}
\bibinfo{author}{D.~M. Newns}, \bibinfo{journal}{Phys.\ Rev.}
  \bibinfo{volume}{\textbf{178}}, \bibinfo{pages}{1123} (\bibinfo{date}{1969}).
\bibitem{:Roh2007}
\bibinfo{author}{M.~Rohlfing}, \bibinfo{author}{R.~Temirov}, and
  \bibinfo{author}{F.~S. Tautz}, \bibinfo{journal}{Phys.\ Rev.\ B}
  \bibinfo{volume}{\textbf{76}}, \bibinfo{pages}{115421}
  (\bibinfo{date}{2007}).
\bibitem{:Kil2008}
\bibinfo{author}{L.~Kilian}, \bibinfo{author}{A.~Hauschild},
  \bibinfo{author}{R.~Temirov}, \bibinfo{author}{S.~Soubatch},
  \bibinfo{author}{A.~Sch\"oll}, \bibinfo{author}{A.~Bendounan},
  \bibinfo{author}{R.~Reinert}, \bibinfo{author}{F.~S. Tautz},
  \bibinfo{author}{M.~Sokolowski}, and \bibinfo{author}{E.~Umbach},
  \bibinfo{journal}{Phys.\ Rev.\ Lett.} \bibinfo{volume}{\textbf{100}},
  \bibinfo{pages}{136103} (\bibinfo{date}{2008}).
\bibitem{:Li2006}
\bibinfo{author}{Z.~Li}, \bibinfo{author}{B.~Han},
  \bibinfo{author}{G.~Meszaros}, \bibinfo{author}{I.~Pobelov},
  \bibinfo{author}{T.~Wandlowski}, \bibinfo{author}{A.~Blaszczyk}, and
  \bibinfo{author}{M.~Mayor}, \bibinfo{journal}{Faraday Discussions}
  \bibinfo{volume}{\textbf{131}}, \bibinfo{pages}{121} (\bibinfo{date}{2006}).
\bibitem{:Hen2007}
\bibinfo{author}{S.~K.~M. Henze}, \bibinfo{author}{O.~Bauer},
  \bibinfo{author}{T.~L. Lee}, \bibinfo{author}{M.~Sokolowski}, and
  \bibinfo{author}{F.~S. Tautz}, \bibinfo{journal}{Surf. Sci.}
  \bibinfo{volume}{\textbf{601}}, \bibinfo{pages}{1566} (\bibinfo{date}{2007}).
\bibitem{:Ere2004}
\bibinfo{author}{M.~Eremtchenko}, \bibinfo{author}{D.~Bauer},
  \bibinfo{author}{J.~A. Sch\"afer}, and \bibinfo{author}{F.~S. Tautz},
  \bibinfo{journal}{New J. Phys.} \bibinfo{volume}{\textbf{6}},
  \bibinfo{pages}{4} (\bibinfo{date}{2004}).
\bibitem{:And61}
\bibinfo{author}{P.~W. Anderson}, \bibinfo{journal}{Phys.\ Rev.}
  \bibinfo{volume}{\textbf{124}}, \bibinfo{pages}{41} (\bibinfo{date}{1961}).
\bibitem{:Kri80}
\bibinfo{author}{H.~R. Krishna-Murthy}, \bibinfo{author}{J.~W. Wilkins}, and
  \bibinfo{author}{K.~G. Wilson}, \bibinfo{journal}{Phys.\ Rev.\ B}
  \bibinfo{volume}{\textbf{21}}, \bibinfo{pages}{1003} (\bibinfo{date}{1980}).
\bibitem{:Pet2006}
\bibinfo{author}{R.~Peters}, \bibinfo{author}{T.~Pruschke}, and
  \bibinfo{author}{F.~B. Anders}, \bibinfo{journal}{Phys.\ Rev.\ B}
  \bibinfo{volume}{\textbf{74}}, \bibinfo{pages}{245114}
  (\bibinfo{date}{2006}).
\bibitem{:Wei2007}
\bibinfo{author}{A.~Weichselbaum} and \bibinfo{author}{J.~v.~Delft},
  \bibinfo{journal}{Phys.\ Rev.\ Lett.} \bibinfo{volume}{\textbf{99}},
  \bibinfo{pages}{076402} (\bibinfo{date}{2007}).
\bibitem{:Kra2006a}
\bibinfo{author}{A.~Kraft}, \bibinfo{author}{R.~Temirov},
  \bibinfo{author}{S.~K.~M. Henze}, \bibinfo{author}{S.~Soubatch},
  \bibinfo{author}{M.~Rohlfing}, and \bibinfo{author}{F.~S. Tautz},
  \bibinfo{journal}{Physical Review B} \bibinfo{volume}{\textbf{74}},
  \bibinfo{pages}{041402} (\bibinfo{date}{2006}).
\bibitem{:Par2006}
\bibinfo{author}{J.~J. Parks}, \bibinfo{author}{A.~R. Champagne},
  \bibinfo{author}{G.~R. Hutchison}, \bibinfo{author}{S.~Flores-Torres},
  \bibinfo{author}{H.~D. {Abru{\~{n}}a}}, and \bibinfo{author}{D.~C. Ralph},
  \bibinfo{journal}{Phys.\ Rev.\ Lett.} \bibinfo{volume}{\textbf{99}},
  \bibinfo{pages}{026601} (\bibinfo{date}{2007}).
\bibitem{SIESTA}
For details about the SIESTA package please refer to
  http://www.uam.es/departamentos/ciencias/fismateriac/.
\bibitem{SIESTA1}
\bibinfo{author}{P.~Ordejon}, \bibinfo{author}{E.~Artacho}, and
  \bibinfo{author}{J.~M. Soler}, \bibinfo{journal}{Phys.\ Rev.\ B}
  \bibinfo{volume}{\textbf{53}}, \bibinfo{pages}{R10441}
  (\bibinfo{date}{1996}).
\bibitem{SIESTA2}
\bibinfo{author}{J.~M. Soler}, \bibinfo{author}{E.~Artacho},
  \bibinfo{author}{J.~D. Gale}, \bibinfo{author}{A.~Garcia},
  \bibinfo{author}{J.~Junquera}, \bibinfo{author}{P.~Ordejon}, and
  \bibinfo{author}{D.~Sanchez}, \bibinfo{journal}{J.\ Phys.-Condens.\ Matter}
  \bibinfo{volume}{\textbf{14}}, \bibinfo{pages}{2745} (\bibinfo{date}{2002}).
\bibitem{:Hau2005}
\bibinfo{author}{A.~Hauschild}, \bibinfo{author}{K.~Karki},
  \bibinfo{author}{B.~C.~C. Cowie}, \bibinfo{author}{M.~Rohlfing},
  \bibinfo{author}{F.~S. Tautz}, and \bibinfo{author}{M.~Sokolowski},
  \bibinfo{journal}{Phys.\ Rev.\ Lett.} \bibinfo{volume}{\textbf{94}},
  \bibinfo{pages}{036106} (\bibinfo{date}{2005}).
\bibitem{:Guti2005}
\bibinfo{author}{R.~Guti{\'e}rrez}, \bibinfo{author}{S.~Mandal}, and
  \bibinfo{author}{G.~Cuniberti}, \bibinfo{journal}{Nano Lett.}
  \bibinfo{volume}{\textbf{5}}, \bibinfo{pages}{1093} (\bibinfo{date}{2005}).
\bibitem{:Pump2007}
\bibinfo{author}{F.~Pump} and \bibinfo{author}{G.~Cuniberti},
  \bibinfo{journal}{Surf. Sci.} \bibinfo{volume}{\textbf{601}},
  \bibinfo{pages}{4109} (\bibinfo{date}{2007}).
\bibitem{:Song2007}
\bibinfo{author}{B.~Song}, \bibinfo{author}{D.~A. Ryndyk}, and
  \bibinfo{author}{G.~Cuniberti}, \bibinfo{journal}{Phys. Rev. B}
  \bibinfo{volume}{\textbf{76}}, \bibinfo{pages}{045408}
  (\bibinfo{date}{2007}).
\bibitem{:Ryn2008}
\bibinfo{author}{D.~A. Ryndyk}, \bibinfo{author}{P.~D'Amico},
  \bibinfo{author}{G.~Cuniberti}, and \bibinfo{author}{K.~Richter},
  \bibinfo{journal}{Phys. Rev. B} \bibinfo{volume}{\textbf{78}},
  \bibinfo{pages}{085409} (\bibinfo{date}{2008}).
\bibitem{gDFTB1}
\bibinfo{author}{A.~Pecchia} and \bibinfo{author}{A.~{Di~Carlo}},
  \bibinfo{journal}{Rep. Prog. Phys.} \bibinfo{volume}{\textbf{67}},
  \bibinfo{pages}{1497} (\bibinfo{date}{2004}).
\bibitem{gDFTB2}
\bibinfo{author}{A.~{Di~Carlo}}, \bibinfo{author}{A.~Pecchia},
  \bibinfo{author}{L.~Latessa}, \bibinfo{author}{T.~Frauenheim}, and
  \bibinfo{author}{G.~Seifert}, \bibinfo{journal}{Chapter 5 in \cite{:Cun2005}}
  .
\bibitem{DFTB1}
\bibinfo{author}{M.~Elstner}, \bibinfo{author}{D.~Porezag},
  \bibinfo{author}{J.~Elsner}, \bibinfo{author}{G.~Jungnickel},
  \bibinfo{author}{M.~Haugk}, \bibinfo{author}{T.~Frauenheim},
  \bibinfo{author}{S.~Suhai}, and \bibinfo{author}{G.~Seifert},
  \bibinfo{journal}{Phys.\ Rev.\ B} \bibinfo{volume}{\textbf{58}},
  \bibinfo{pages}{7620} (\bibinfo{date}{1998}).
\bibitem{:Szu2003}
\bibinfo{author}{B.~Szuecs}, \bibinfo{author}{Z.~Hajnal},
  \bibinfo{author}{T.~Frauenheim}, \bibinfo{author}{C.~Gonz{\'a}lez},
  \bibinfo{author}{J.~Ortega}, \bibinfo{author}{R.~P{\'e}rez}, and
  \bibinfo{author}{F.~Flores}, \bibinfo{journal}{Appl. Surf. Sci.}
  \bibinfo{volume}{\textbf{212}}, \bibinfo{pages}{861} (\bibinfo{date}{2003}).
\bibitem{:Szu2004}
\bibinfo{author}{B.~Szuecs}, \bibinfo{author}{Z.~Hajnal},
  \bibinfo{author}{R.~Scholz}, \bibinfo{author}{S.~Sanna}, and
  \bibinfo{author}{T.~Frauenheim}, \bibinfo{journal}{Appl. Surf. Sci.}
  \bibinfo{volume}{\textbf{234}}, \bibinfo{pages}{173} (\bibinfo{date}{2004}).
\bibitem{DFTB}
Please refer to http://www.dftb.org.

\end{thebibliography}

\end{document}